\documentclass[authoryear]{elsarticle}

\usepackage{amsmath,amssymb,graphicx,pifont}
\usepackage{lineno}
\usepackage[usenames]{color}
\definecolor{lightgray}{gray}{0.2}

\renewcommand{\d}{\text{d}}
\newcommand{\degr}{$^{\circ}$} 
\begin{document}

\title{Oscillatory thermal instability --- the Bhopal disaster and liquid bombs}

\author{Rowena Ball}
\ead{Rowena.Ball@anu.edu.au}

\address{Mathematical Sciences Institute, The Australian National University\\
 Canberra ACT 0200 Australia
}

\begin{abstract}
Thermal runaway reactions were involved in the Bhopal disaster of 1984, in which methyl isocyanate was vented from a storage tank of the liquid, and occur in liquid peroxide explosions,  yet  to date there have been few investigations into the mechanism of thermal runaway in such liquid thermoreactive systems. Consequently protocols for  storing thermally unstable liquids and deactivating liquid bombs may be suboptimal. In this work the hydrolysis of methyl isocyanate and the thermal decomposition of triacetone triperoxide were simulated using a gradientless, continuous-flow reactor paradigm. This approximation enabled stability analyses on the steady state solutions of the dynamical mass and enthalpy equations. The results indicate that thermal runaway in both systems  is due to the onset of a large amplitude, hard thermal oscillation initiated at a subcritical Hopf bifurcation. This type of thermal misbehaviour cannot be predicted using classical ignition theory, and may be typical of liquid thermoreactive systems.  
The  mechanism of oscillatory thermal instability on the nanoscale is elucidated. 
\end{abstract}

\maketitle
\section{Introduction \label{sec1}}

The causes of the world's worst industrial disaster at Bhopal on 2--3 December 1984 are still being debated in the international media, more than 25 years after the event. Was it caused by neglect, parsimony, or procrastination by Union Carbide on training, safety and maintenance? Corruption, sabotage and cover-up? Inadequate government regulation? Any or all of the above may well have helped set up the worst possible scenario --- for it could not have been any worse --- but they are  contributing factors rather than causes. (A brief account of the disaster is given in the Appendix.) 

The primary \textit{cause} of the thermal runaway that led to the venting of a lethal mist of methyl isocyanate  (MIC) over Bhopal city was physicochemical.  In this work I present a stability analysis of a simple dynamical model for the MIC-H$_2$O reacting system, revealing oscillatory thermal misbehaviour that cannot be predicted using classical ignition theory.  A similar instability is shown to govern the explosive thermal decomposition of the organic hydroperoxide triacetone triperoxide, in liquid solution. The provenance of oscillatory thermal instability on the nanoscale is elucidated, and shown to lie in the ability of the reactant molecules to store energy in the internal molecular motions --- in other words, the heat capacity. 

Despite the enormity of the Bhopal disaster little or no research has been published that elucidates the fundamental physico-chemical cause of thermal runaway in liquid reactive systems such as MIC hydrolysis. In terms of achieving the Millenium Development Goals (MDG) it seems rather important that thermoreactive processes in liquids are thoroughly investigated and the knowledge disseminated widely.  Given the the horrific legacy of the disaster, the long term adverse health effects in children and adverse reproductive effects in women of MIC exposure that have been well-documented \citep{Mishra:2009},  such knowledge is relevant to the MDG of Child Health and Maternal Health.  More generally people have a right to expect that thermally unstable and hazardous liquids are stored safely. 

The~Millennium~Development Goals~Report~\citep{mdg:2010} highlights the challenges posed by conflicts and armed violence to human security and MDG achievements. A new and growing threat to people's security is the use of liquid peroxide explosives by terrorists. The ingredients for making such bombs are cheap and widely available and they cannot be detected by metal detectors and nitrogenous explosives detectors, or distinguished from hand lotion by x-ray machines. Liquid peroxide-based explosives were used in the suicide attacks on the London transit system in 2005, which killed 56 people, and  the terrorists convicted of the foiled 2006 transatlantic aircraft conspiracy had plotted to blow up a number of planes using liquid peroxide explosives. (Many more accounts of  peroxide misuse incidents are easily found on the web.) To this day there are severe restrictions on carrying liquids through security barriers at most airports. It seems grimly inevitable that the use of liquid peroxide explosives as mass murder weapons will increase. Knowledge of their fundamental mechanism of action may help to counter their use.



The open literature on theoretical and experimental validations of thermal  runaway criteria and parametric sensitivity  in batch reactors and storage tanks was summarized by  \citet{Velo:1996}. In defining critical conditions they, along with  other authors cited therein, begin with the assumption that storage tanks can be modelled as well-stirred batch reactors with linear thermal coupling to the environment. 
However batch reactors have no non-trivial steady states, and there is no general theory for determining whether a thermal excursion will grow or decay.  It is  shown in this work that a simple model with nonequilibrium steady states that is also spatially homogeneous  --- the continuous-flow stirred tank reactor (CSTR) paradigm --- can provide great insight into thermoreactive instabilities in liquid systems, and provide fundamental causative information that cannot easily be extracted from detailed numerical simulations that include convective motions. 

In section \ref{sec2} I describe the chemical reactions and provide the relevant data, taken from the literature, for the physical properties of the reactants and thermodynamic and kinetic parameters. The CSTR paradigm is described in section \ref{sec3} and the equations are given using dimensionally consistent units and in terms of dimensionless variables and parameter groups. In section \ref{sec4} the results of numerical stability analyses of the equations are given, where numerical values of the parameters for MIC hydrolysis and for triacetone triperoxide thermal decomposition in solution were used in the equations. Some points regarding the applicability of the CSTR paradigm are discussed in section \ref{sec5}, and the nanoscale aspects of oscillatory thermal runaway are elucidated through examining the behaviour in the limits of the two timescales of the relaxation oscillation. A summary of the conclusions is given in section \ref{sec6}.

\section{ \label{sec2}Chemistry and  data} 

\subsection{MIC hydrolysis}

Isocyanates hydrolyse exothermically to the corresponding amine and carbon dioxide.  In excess water isocyanates react exothermically with the hydrolysis product amine to form the disubstituted urea \citep{Saunders:1948,Dsilva:1986}. With MIC the product is N,N-dimethyl urea and the reaction sequence is as follows:
 \begin{align} 
 \text{CH}_3\text{NCO}_\text{(l)}  + \text{H}_2\text{O}_\text{(l)}  &\overset{k_1(T)}{\longrightarrow}  \text{CH}_3\text{NH}_{2\text{(aq)}}  + \text{CO}_{2\text{(aq)}}  \tag{R1}\\
 \text{CH}_3\text{NCO}_\text{(l)} + \text{CH}_3\text{NH}_{2\text{(aq)}}  &\overset{k_2(T)}{\longrightarrow}  \text{CH}_3\text{NHCONHCH}_{3\text{(aq)}} .\tag{R2}
 \end{align} 
 For reaction R2 $\Delta H_2 (298\,\text{K})= -174.6$\,kJ/mol and for the sequence overall $\Delta H_\text{tot}(298\,\text{K})=-255$\,kJ/mol \citep{Lide:2009}. 
 

A chemical analysis of the residue in the MIC storage tank (Tank 610) at the Union Carbide plant at Bhopal, sampled seventeen days after the event, found a variety of MIC condensation products, mainly the cyclic trimer \citep{Dsilva:1986}.  However, experiments  indicated that these condensations must have been initiated at temperatures and pressures well above the normal boiling point of MIC, so for modelling the initial thermal runaway these reactions need not be considered. No kinetic data are available for reaction R2 so only reaction R1 is used in the model. It will be seen from the results that reaction R1 alone is sufficient to induce thermal runaway. Relevant physicochemical data are given in table \ref{table1}. 
\begin{table}\caption{\label{table1}Physical, kinetic, and thermochemical parameters for MIC hydrolysis.}
{\begin{tabular}{p{0.55\columnwidth}p{0.15\columnwidth}p{0.2\columnwidth}}
\hline\hline
Molecular weight MIC &57.051&\\
Specific heat capacity $C_p^\circ$(298) MIC &1959\,J/(kg\,K)& \citet{Perry:2008}\\
Specific heat capacity $C_p^\circ$(298) H$_2$O&4181\,J/(kg\,K)&\\
Boiling point  MIC at 1\,atm& 38.3{\degr}C& \citet{Lide:2009}\\
Density  MIC  at 25{\degr}C & 0.9588\,g/cm$^3$&\citet{Lide:2009} \\
R1 reaction enthalpy& 80.4\,kJ/mol&\citet{Lide:2009}$^\dag$\\
R1 activation energy&65.4\,kJ/mol&\citet{Castro:1985}\\
R1 pseudo first order frequency factor & 4.13e08/s&\citet{Castro:1985}\\
\hline\hline
\end{tabular}}
$^\dag$From standard enthalpies of formation at 298\,K. 
\end{table}

\subsection{Thermal decomposition of triacetone triperoxide (TATP)}
Triacetone triperoxide, a cycle trimer,  is an explosive made by mixing acetone and hydrogen peroxide, both of which substances are legal, cheap and readily available over the counter. Pure TATP is a white crystalline powder that is soluble in organic solvents. The thermal decomposition of TATP does not involve combustion; the main reaction products are acetone,  some carbon dioxide, and ozone \citep{Eyler:2000,Oxley:2002}. Its high explosive power is in part due to the large entropy increase of the formation of four gas molecules from one condensed-phase molecule \citep{Dubnikova:2005}. Relevant parameters for the thermal decomposition of TATP in toluene are given in table \ref{tatp}. 

\begin{table}\caption{\label{tatp}Physical, kinetic, and thermochemical parameters for thermal decomposition of TATP in toluene.}
{\begin{tabular}{p{0.55\columnwidth}p{0.15\columnwidth}p{0.2\columnwidth}}
\hline\hline
Molecular weight  &222.2356\,g/mol&\\
Specific heat capacity $C_p^\circ$(298) toluene &1698.25\,J/(kg\,K)& \citet{Perry:2008}\\
Boiling point toluene  at 1\,atm&110.8 {\degr}C& \\
Density of toluene at 298\,K&866.9\,kg/m$^3$&\\
Reaction enthalpy & 330--420\,kJ/mol$^\dag$&\cite{Dubnikova:2005}\\
Activation energy&178.52\,kJ/mol&\citet{Eyler:2000}\\
Frequency factor & 9.57e16/s&\citet{Eyler:2000}\\
Feed concentration of TATP& 2\,mol/kg&\\
\hline\hline
\end{tabular}}
$^\dag$Depending on reaction products. 
\end{table}

\section{The CSTR paradigm\label{sec3}}
 The spatially homogeneous flow reactor, or reacting mass or volume, in which a reactant undergoes a first order, exothermic conversion is a simple but elucidatory model for thermoreactive systems when it is appropriate to ignore convection, because as a dynamical system it has non-trivial steady states that can be analysed for stability. The dynamical mass and enthalpy equations may be written as
\begin{align}
M\frac{\text{d}c}{\text{d}t} =  &Mze^{-E/RT}c + F(c_f-c) \label{e1}\\
MC_r\frac{\text{d}T}{\text{d}t} &=  (-\Delta H)Mze^{-E/RT}c +F(C_fT_a-C_rT) -L(T-T_a)  \label{e2}.
\end{align}
The symbols and quantities are defined in table \ref{table2}. 
For numerical and comparative reasons  it is more convenient to work with the following dimensionless system corresponding to equations  (\ref{e1}--\ref{e2}):
\begin{align}
\frac{\text{d}x}{\text{d}\tau}&=-xe^{-1/u}+f(1-x)\label{e4}\\
\varepsilon\frac{\text{d}u}{\text{d}\tau}&=xe^{-1/u} +\varepsilon f(\gamma u_a- u) - \ell(u-u_a),\label{e5}
\end{align}
where the dimensionless groups are defined in table \ref{table2}.
Numerical analysis of equations (\ref{e4}--\ref{e5}) was carried out using values of the dimensionless groups calculated from the data in tables \ref{table1} and table \ref{tatp} and assigned values of the inverse residence time $f$, heat loss coefficient $\ell$, and inflow concentration $c_f$.  

\begin{table}[h]\caption{\label{table2}Quantities,  definitions,   and units.}
{\vbox{\begin{tabular}{p{0.04\columnwidth}p{0.5\columnwidth}p{0.04\columnwidth}p{0.1\columnwidth}}
\hline\hline
$A$& reaction frequency&&s$^{-1}$\\
$c$ & $c(t)$, concentration of reactant & &mol/kg\\
$c$ & inflow reactant concentration & &mol/kg\\
$C_r$&specific heat capacity of reaction mixture& &J/kg\,K\\
$C_f$&specific heat capacity of the inflow stream& &J/kg\,K\\
$E$ & activation energy &&J/mol\\
$F$ & flow through rate  & &kg/s\\
$\Delta H$& reaction enthalpy& &J/mol\\
$M$ &mass of reaction mixture & &kg\\
$R$ &gas constant & 8.314&J/mol\,K\\
$t$ & time & &s\\
$T$& $T(t)$ reaction temperature &&K\\
$T_a$ & ambient temperature &&K\\
$L$ & heat loss coefficient &&W/K\\ 
\hline
\end{tabular}
\begin{tabular}{p{0.04\columnwidth}p{0.4\columnwidth}p{0.04\columnwidth}p{0.2\columnwidth}}
$\varepsilon$&${C}_rE/c_f(-\Delta H)R$& $\tau$& $ tA$\\
$f$&$ F/MA$&$u$&$ RT/E$\\
$\gamma$&$C_f/C_r$&$u_a$&$ RT_a/E$\\
$\ell$&$ LE/c_fMA(-\Delta H)R$&$x$ & $ c/c_f$\\ 
\hline\hline
\end{tabular}}
}
\end{table}

\section{Results} \label{sec4}
\subsection{MIC hydrolysis}

\begin{figure}[ht]
\centerline{
\includegraphics[scale=0.6]{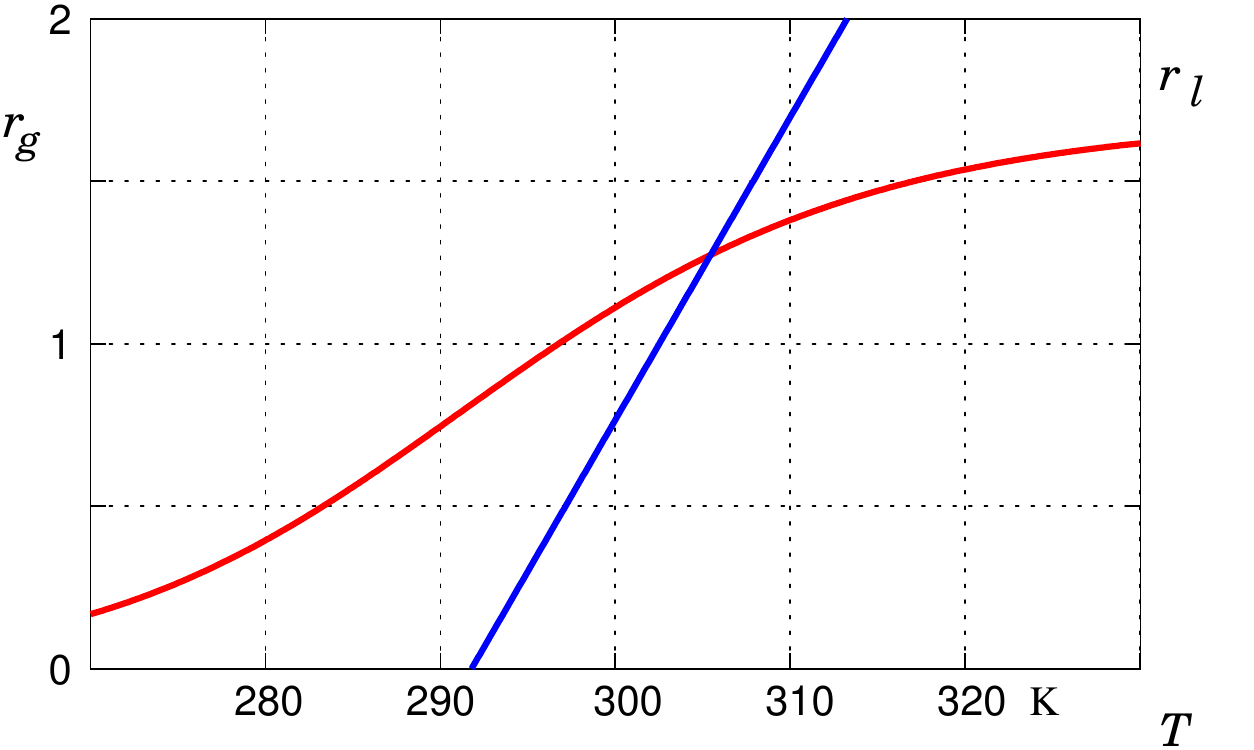}}
  \caption[]{\label{figure1}Rates of reactive heat generation $r_g$ (red) and heat loss $r_l$ (blue) versus  $T$ from equations (\ref{e4}--\ref{e5}). $u_a=0.0379$ (corresponding to $T_a=292\,K $), $f=1.7$, $\ell=700$, $\varepsilon=10$. }
\end{figure}

From equations \ref{e4} and \ref{e5} in the steady state we can define the rate of reactive heat generation as the nonlinear term
$$
r_g\equiv fe^{-1/u}/(e^{-1/u}+f),
$$
and the rate of non-reactive cooling as the linear term
$$
r_c\equiv -u(\varepsilon f+\ell)+u_a(\varepsilon\gamma f+\ell).
$$
From classical ignition theory the reacting mixture self-heats if $r_g$ exceeds $r_l$. Thermal runaway occurs if $r_g$ exceeds $r_l$ beyond a system-specific threshold; for the hydrolysis of MIC this is taken as the boiling point of MIC. These rates were computed using data from table \ref{table1} and  plotted in figure~\ref{figure1}, where the temperature is labeled in dimensional units. 
We see that the system self-heats until the reaction temperature $T$ reaches the steady state temperature of $\sim$305\,K at which the heating and cooling rates are balanced. Since the boiling point of MIC is 312\,K, according to this diagram  the Bhopal disaster did not happen.  On the basis of this diagram we would not expect a thermal runaway to develop, even when the ambient temperature is allowed to drift slowly up to 292\,K. 

\begin{figure}[ht]
\centerline{
\includegraphics[scale=0.9]{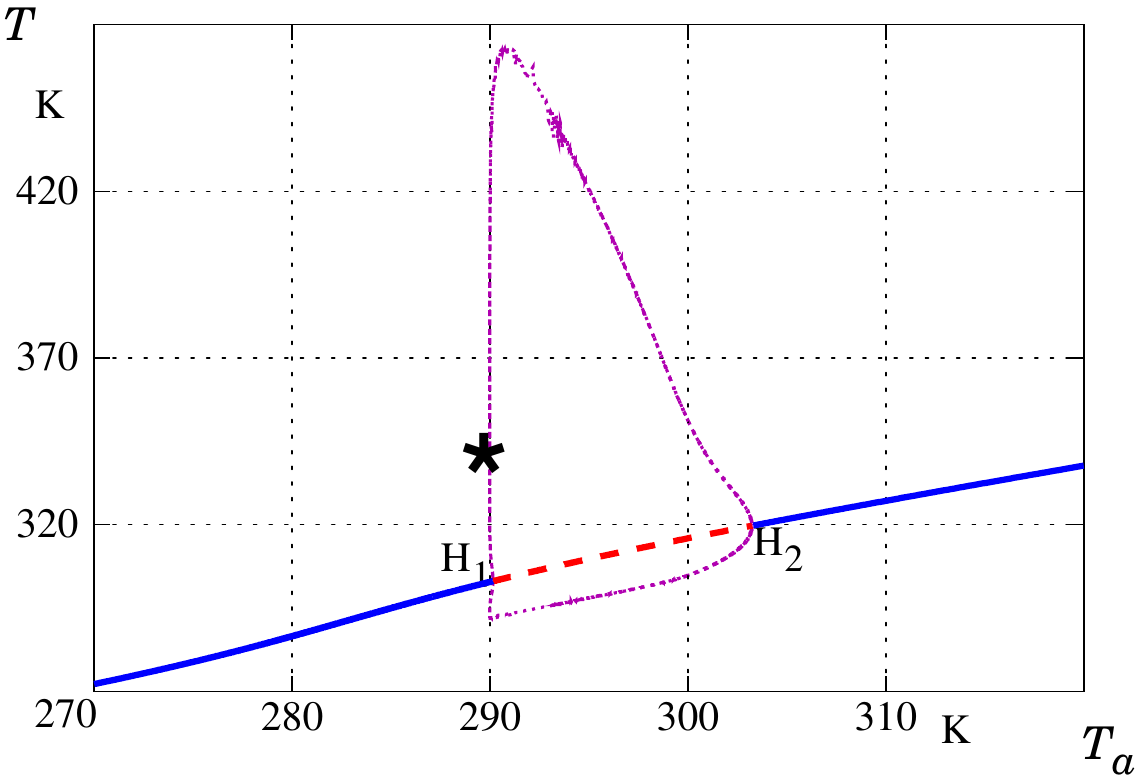} }
  \caption[]{\label{figure2} Bifurcation diagram. Stable steady states are plotted with solid blue line, unstable steady states with dashed red line, and the amplitude envelope of periodic solutions is marked with thin dotted magenta line. $H_1$ and $H_2$ label the Hopf bifurcation and the large  \textbf{\ding{81}} marks the change in stability of the limit cycles. $f=1.7$, $\ell=700$, $\varepsilon=10$.}
\end{figure}

However thermal balance diagrams such as that in figure \ref{figure1} can be dangerously misleading because they infer stability rather than assess stability rigorously, although such diagrams are often used in chemical reactor engineering. The steady states, periodic solutions, and stability analysis of equations (\ref{e4}--\ref{e5}) were computed numerically \citep{Doedel} and yielded a dramatically different picture of the the thermal stability of MIC hydrolysis. Figure \ref{figure2} shows a bifurcation diagram in which the steady state temperature  and the temperature amplitude envelope of periodic solutions are plotted as a function of $T_a$. 

The steady state is stable at $T_a\approx 286\,K$, the temperature at which the tank of MIC had been held for several months. As $T_a$ is quasistatically increased the reaction temperature $T$ increases slowly, but at $T_a=290.15\,K$ the stability analysis flags an abrupt change in the nature of the solutions. At this point the steady state solutions become unstable at a Hopf bifurcation and the hard onset of a high amplitude thermal oscillation ensues. Clearly, at $T_a=292\,K $ we have catastrophic thermal runaway, contrary to the  prediction given by figure \ref{figure1}. (In the resulting superheated fluid the exothermic condensation reactions would increase the temperature even further.) 

\begin{figure}[t]
\centerline{
\includegraphics[scale=0.7]{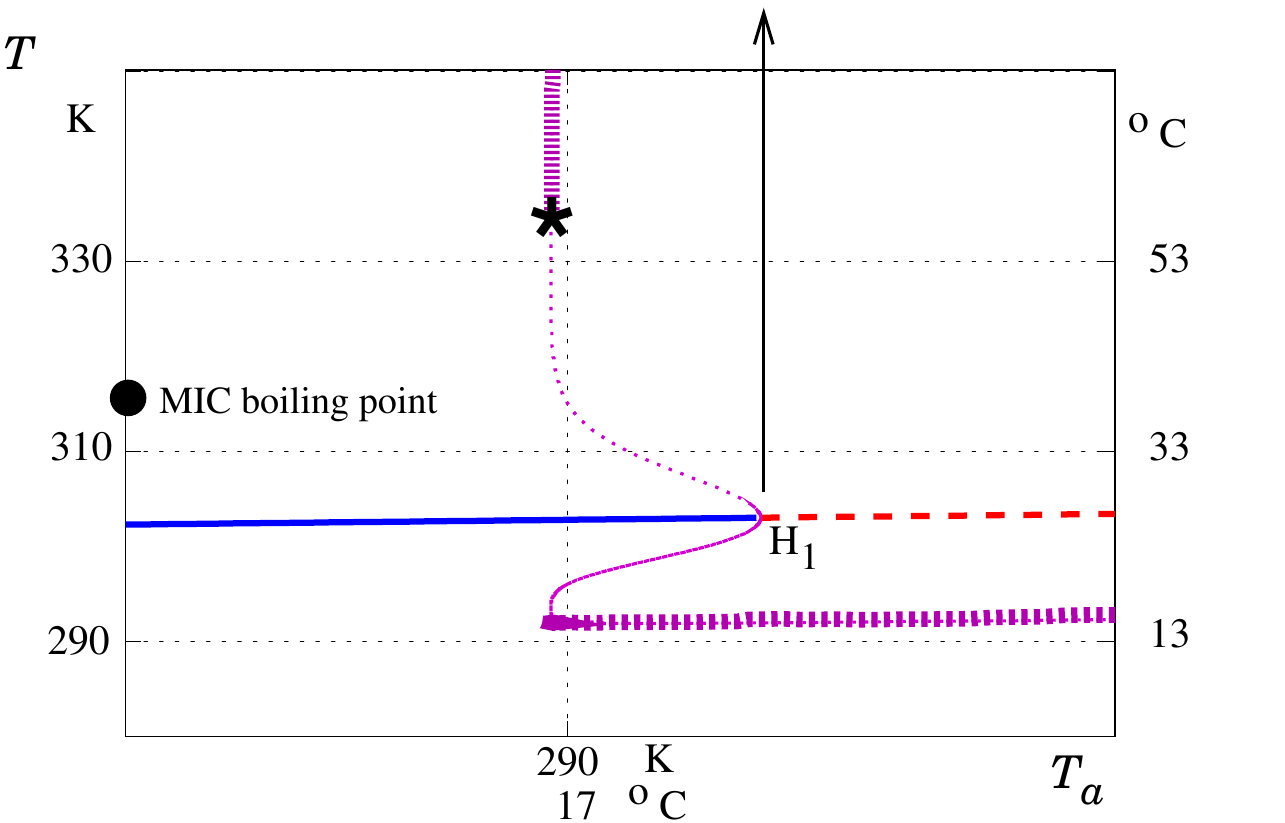}}
  \caption[]{\label{figure3} Close-up of the region around the Hopf bifurcation $H_1$ in figure \ref{figure2}. }
\end{figure}

This is quite different from classical ignition of a thermoreactive system, which occurs at a steady-state turning point. The dynamics of oscillatory thermal runaway can be understood by studying the close-up of the region around the Hopf bifurcation $H_1$ shown in figure  \ref{figure3}. 
$H_1$ is subcritical and the emergent limit cycle is \textbf{un}stable. The amplitude envelope of the unstable limit cycles is marked with a thin dotted line; they grow as $T_a$ is \textbf{de}creased. At the turning point \raisebox{-1ex}{\Large\textbf{*}} of the periodic solution branch  the limit cycles become stable. Thermal runaway \textit{may} occur if there are significant perturbations while $T_a$ is within the regime \raisebox{-1ex}{\Large\textbf{*}}--$H_1$, and it \textit{must} occur when $T_a$ drifts above $H_1$.  In principle the rapid thermal excursion takes the system to the amplitude maximum of the stable limit cycle.  In reality the reactant and products have vaporised, the pressure has soared,  the peak temperature is far above the boiling point of MIC, and the system must vent or explode. But it must be emphasized that the thermal runaway is due to oscillatory instability rather than classical ignition at a turning point.

\begin{figure}[h]
\centerline{
\includegraphics[scale=0.7]{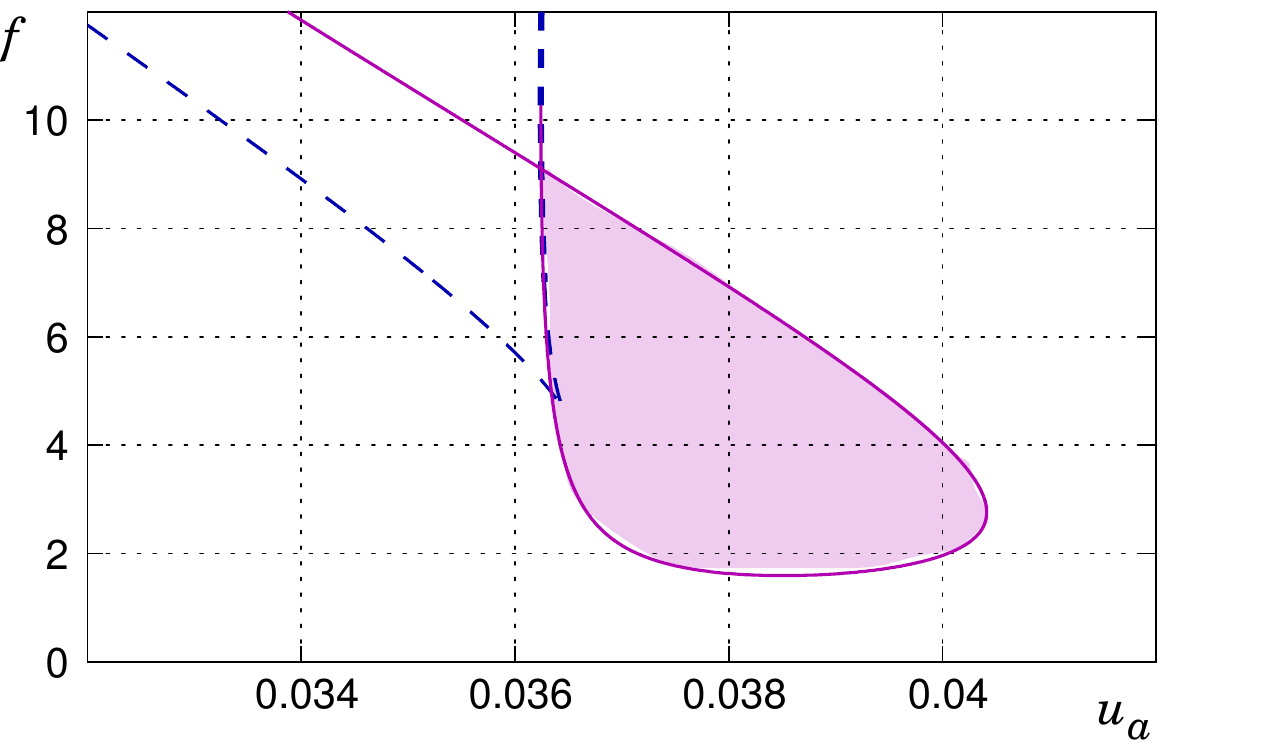}}
  \caption[]{\label{figure4} The locus of Hopf bifurcations is marked with a solid line, the locus of steady-state turning points is marked with dashed line.  $\ell=700$, $\varepsilon=10$. }
  \end{figure}

The presence of oscillatory instability is all-pervasive and dominant in this system. This can be appreciated by inspection of figure \ref{figure4}, 
  a plot of the loci of the steady state turning points and the Hopf bifurcations of equations (\ref{e4}--\ref{e5}) over the two parameters $u_a$ and the inverse residence time $f$. 
In the filled region thermal runaway will always be oscillatory. The bistable regime, indicated by the dashed line, occurs at very high flow rates (short residence times). However, classical thermal runaway at a steady state turning point does not occur because the oscillatory instability is still present and dominant.

\begin{figure}[h]
\centerline{
\includegraphics[scale=.4]{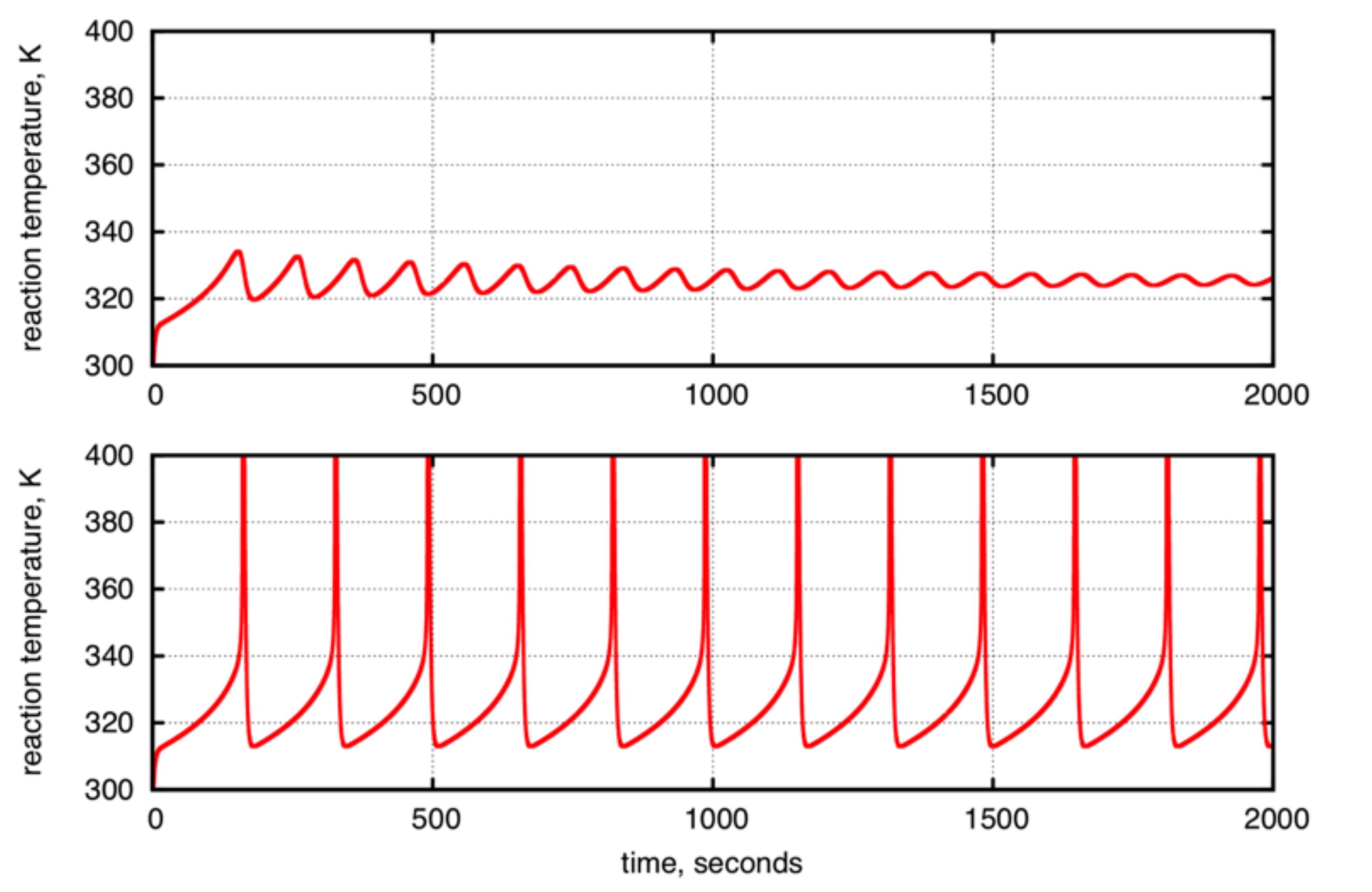}
}
\caption{\label{figure5} Computed time series for MIC hydrolysis with $F=0.0016$\,kg/s,  $L=560$\,W/K. Upper plot: $T_c=308.4$\,K, lower plot: $T_c=308.5$\,K}
\end{figure}

Two computed time series for $F=0.0016$\,kg/s are compared in figure \ref{figure5}. 
In the upper plot $T_a=308.4$\,K and the oscillations decay to a stable steady state. However, the onset of thermal instability is violent: in the lower plot $T_a=308.5$\,K and the transient does not decay but swings into sustained high amplitude relaxation oscillations with a period of about 166\,s.

\begin{figure}[h]
\centerline{
\includegraphics[scale=.4]{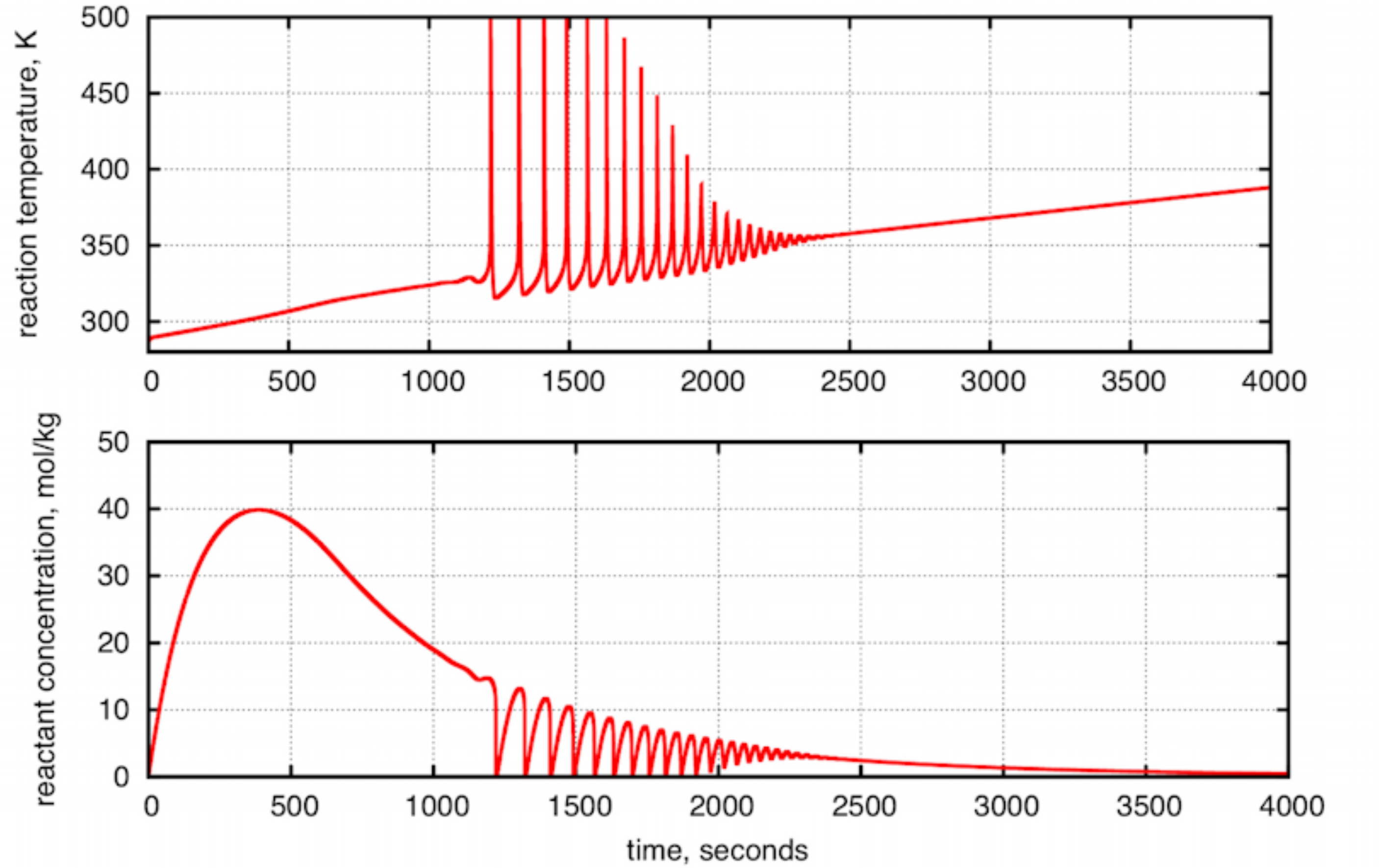}
}
\caption{\label{figure6} Computed time series for MIC hydrolysis with drift in thermostat temperature of 0.02\degr{C}/s.  $F=0.0016$\,kg/s,  $L=560$\,W/K. }
\end{figure}

The behaviour of the system under a slow upwards drift of the ambient temperature can be simulated easily; a time series with $\d T_a/\d t=0.02^\circ{C}$/s is shown in figure \ref{figure6}, 
which confirms the abrupt onset of the instability. Of note is the  decay in amplitude of the oscillations as the thermostat temperature \textit{increases}; physically this occurs because the reactant is consumed faster than it is supplied as the temperature increases. 

\subsection{Triacetone triperoxide}
\begin{figure}[h]
\centerline{
\includegraphics[scale=0.35]{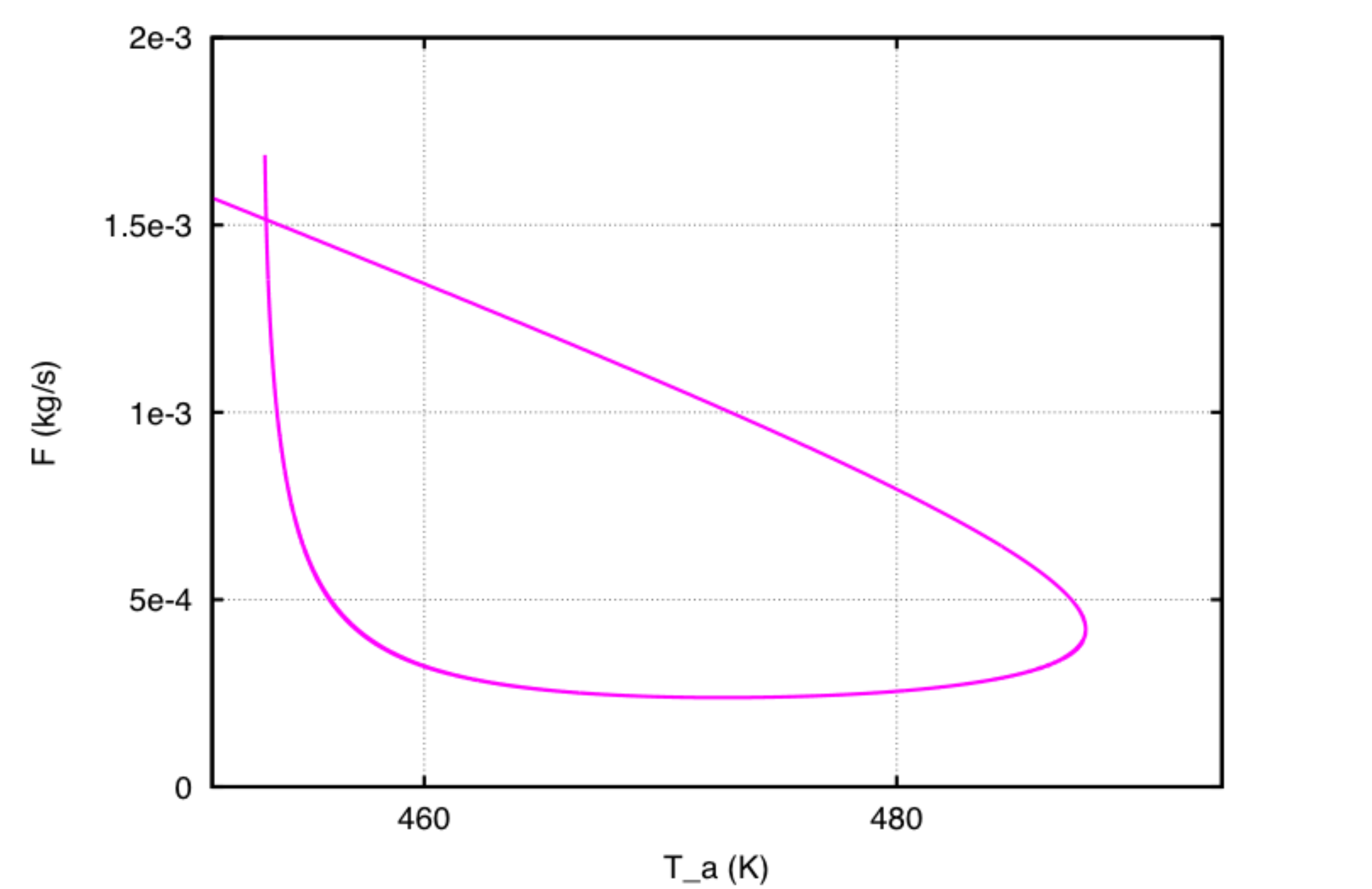}}
  \caption{\label{tatp-2p}  The locus of Hopf bifurcations for the triacetone triperoxide system. }
\end{figure}

\begin{figure}
\centerline{
\includegraphics[scale=0.5]{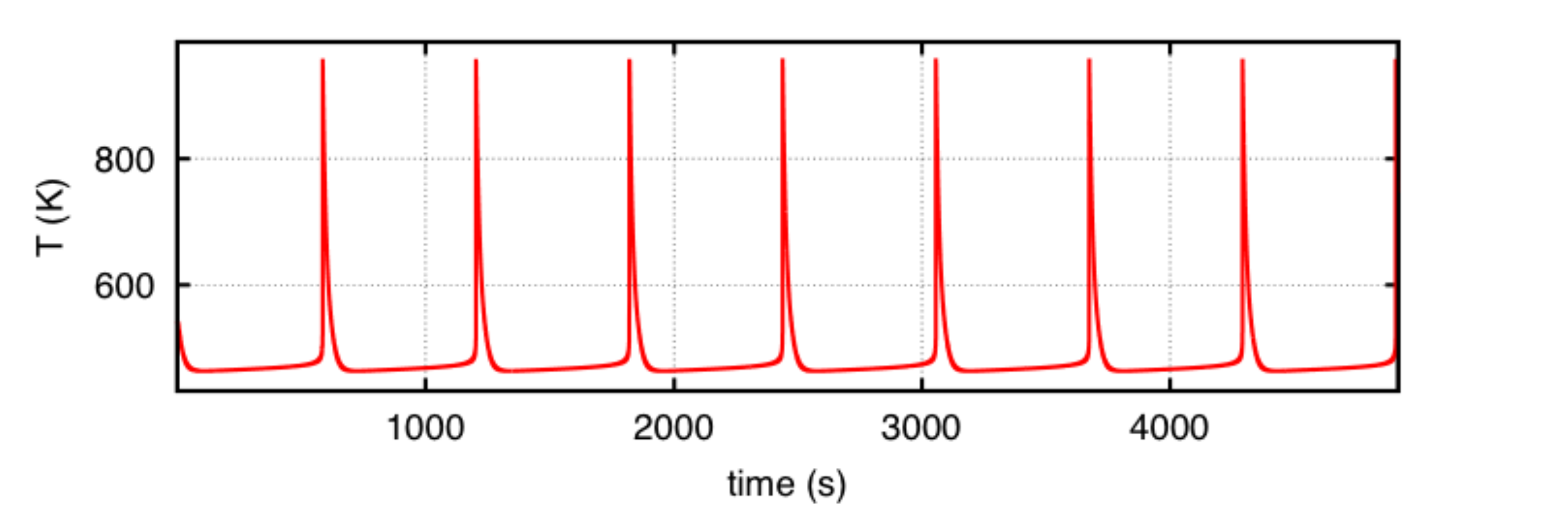}}
  \caption{\label{tatp-ts1} Strong relaxation oscillations in the TATP thermal decomposition. The time series was computed for $T_a=460$\,K, $F=5$e-04\,kg/s. }
\end{figure}

The steady states, periodic solutions, and stability analysis of equations (\ref{e4}--\ref{e5}) were computed using the data for TATP thermal decomposition in table \ref{tatp} and results are shown in figures  \ref{tatp-2p} and \ref{tatp-ts1}.  

In figure \ref{tatp-2p} the loci of the Hopf bifurcations are plotted over $T_a$ and $F$, and a point within the oscillatory region was selected to compute the time series in figure\ref{tatp-ts1}. The system exhibits violent relaxation oscillations, suggesting that explosive thermal decomposition of TATP is initiated at the onset of this oscillatory behaviour rather than by classical ignition. 

\clearpage

\section{Discussion \label{sec5}}

The tendency for oscillatory thermal runaway may be typical of exothermically reactive organic liquids. In the work of \citet{Ball:1995a} the hydration of 2,3-epoxy-1-propanol in a CSTR was found to exhibit similar non-classical thermal misbehaviour. Here it is shown that  the thermal runaway that led to the Bhopal disaster may have been initiated at an oscillatory instability, and that liquid peroxide explosions may be initiated at an oscillatory instability rather than by classical thermal ignition. Similar results have been obtained using parameters for the decomposition of cumene hydroperoxide  in equations (\ref{e4}--\ref{e5}) \citep{Ball:2010}. 

Is it realistic to model a reacting volume inside a storage tank --- or, for that matter, a peroxide bomb --- as a well-stirred flow reactor? Yes, on a timescale over which the reacting volume remains relatively constant and gradientless relative to the much faster rate of reaction. 
For the purposes of this analysis in which the focus is on the dynamics we can circumscribe a reacting volume in which the spatial gradients are insignificant in comparison to the time evolution, and therefore can be neglected.  In this case the CSTR paradigm is appropriate. If this approximation does not hold, then we are free to reduce the circumscribed volume until it does. There is nothing particularly artificial or manipulative in doing this; it is just a simplest case scenario for which the powerful tools of stability and bifurcation theory can be applied. Much of the heat transfer would be convective rather than conductive, and on convective timescales the approximation does not hold --- but that is for a separate study. 

\subsection{Nanoscale aspects of oscillatory thermal instability} 
Since this is a book about all things nano, it is pertinent to elucidate the mechanism of oscillatory thermal instability on the nanoscale. But before we zoom in to nano spatial scales we need to discuss the characteristics of relaxation oscillators in general and examine the two time scales of the relaxation oscillation solutions of equations \ref{e4} and \ref{e5}. 

Relaxation oscillators are often and easily implemented in electrical and electronic circuits,   but any dissipative dynamical system with nonequilibrium steady states  and  two or more dynamical state variables has the potential to exhibit relaxation oscillations. In general they are not smoothly sinusoidal in form. Instead, relaxation oscillators are characterised by energy dissipation and energy accumulation occurring on different timescales. There may be relatively slow dissipation until the system reaches a threshold state at which the internal energy rapidly and nonlinearly increases, or rapid dissipation   followed by slow but accelerating increase  in internal energy. This latter two-time dynamics occurs with the thermochemical oscillator in  equations \ref{e4} and \ref{e5}. 

\begin{figure}
\centerline{
\includegraphics[scale=.25]{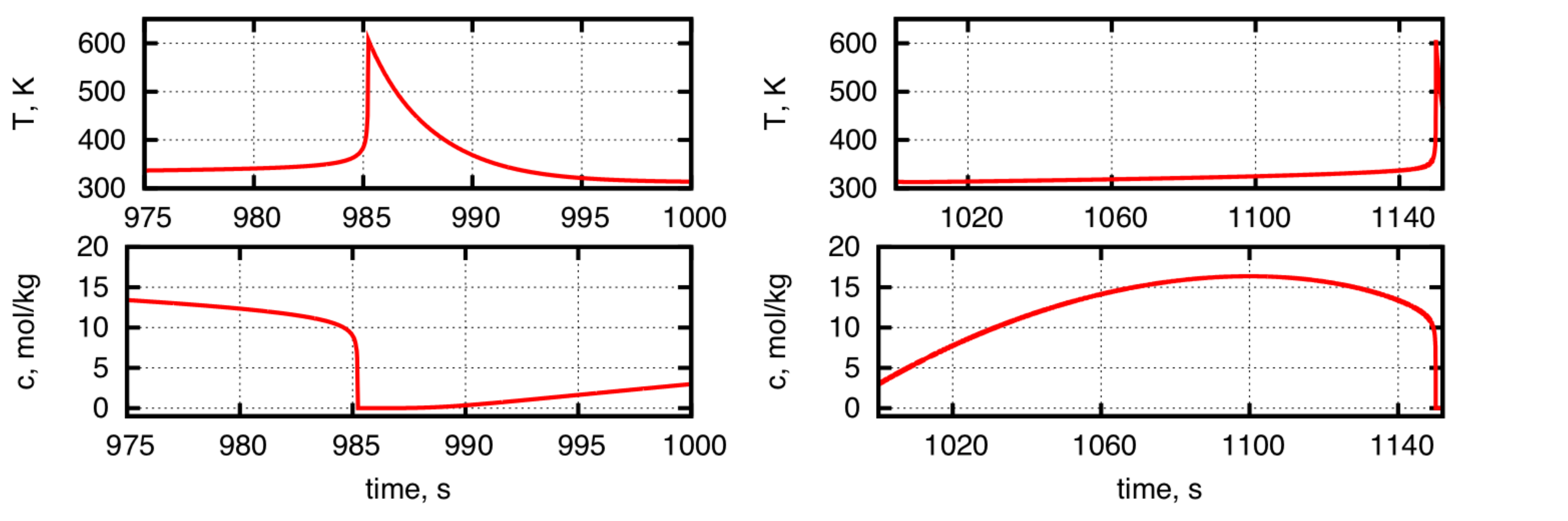}}
  \caption{\label{ro} Left: a `fast' or dissipative time interval. Right: a 'slow' time interval.}
\end{figure}

The two time scales in each period evident in figure \ref{figure5} are shown in figure~\ref{ro}. On the `fast' time scale, left, during the few seconds between the temperature maximum and about 990\,s the concentration remains close to zero. On the `slow' time scale, right, the system barely heats at all over the long interval, although evidently some reaction takes place because reactant accumulates nonlinearly, reaches a maximum, then declines slowly before the exponential spike.  

 We can extract these two time scales approximately  from equations \ref{e4} and \ref{e5}. Define `fast' or `stretched' or dissipative  time  as $\tau^\prime\equiv \tau/\varepsilon $ and recast the equations with  $\tau^\prime$ as the independent variable:
\begin{align}
\frac{\d x}{\d \tau^\prime} &= \varepsilon\left(-xe^{-1/u} + f\left(1-x\right)\right) \label{e6}\\ 
\frac{\d u}{\d  \tau^\prime}&= xe^{-1/u} +\varepsilon f(\gamma u_a- u) - \ell(u-u_a) .\label{e7}
\end{align} 
For $\varepsilon$ sufficiently small we have $\d x/\d  \tau^\prime \approx 0$ and $x\approx x_0 \approx 0$ since the reactant is almost fully depleted, and $\d u/\d \tau^\prime\approx -\ell(u-u_a)$, so that $u$ decays as $u\approx \exp(-\ell\tau^\prime)(u_0 - u_a) + u_a = \exp(-\ell\tau^\prime)(u_\text{max} - u_a) + u_a$ since the temperature amplitude is at a maximum when $x=0$ at $\tau^\prime_\text{max}=\tau^\prime_0\equiv 0$.  
As $\tau^\prime$ becomes `large' $u\rightarrow u_a$; in  figure \ref{ro} left this occurs around $\sim$1000\,s.  However  reactant  is gradually accumulating and the system evolves in  `slow' time. 

Over much of the `slow' time interval the system behaviour can be understood by dividing equation \ref{e5} through by $\varepsilon$.  At low temperatures and with large $\varepsilon$ we then have 
$\d u/\d \tau\approx -f(u-u_a)$, which gives $u\approx \exp(-f\tau)(u_0 - u_a) + u_a$, and since $u_0\equiv u_a $ on this time scale we have $u\approx u_a$. In figure \ref{ro}  this approximation holds over much of the `slow' time interval:  the internal energy of the system increases but the temperature remains almost constant, although slowly increasing. 
 During this time reactant accumulates, but reaction does take place: $x$ evolves as $x\approx f(1-\exp(-b\tau)/b$, where $b=\exp(-1/u_a)$.  

If reaction takes place in `slow' time but the system is barely heating up, where does the heat of reaction go? Now we focus on  the nanoscale. 

The answer is that the heat is stored in the internal motions of the reactant mixture molecules. Referring to the definition of the dimensionless group $\varepsilon$ in table \ref{table2} we see that $\varepsilon$ is large if the specific heat capacity $C_r$ of the reaction mixture is large. The specific heat capacity of a substance is a function of the number of degrees of freedom of motion that are available to its constituent particles. A monoatomic perfect gas has only the kinetic energy of each atom so translational motion in three dimensions is the only motion it can undergo. The equipartition theorem tells us that the three translational contributions to the constant volume molar heat capacity $C_v^\text{tr}=(3/2)R=12.47$\,J/(mol\,K), thus the kinetic energy contribution does manifest as temperature change. 

Polyatomic molecules have many additional degrees of freedom though, and physical properties of  liquids are governed much more by the potential energy of the system than the  kinetic energy.  Potential energy is stored in intramolecular rotational and vibrational degrees of freedom and in the vibrational intermolecular force potential. The mean translational energy of each liquid molecule is the same as that for gases, $(3/2)kT$, where $k$ is Boltzmann's constant, and since $C_p=C_v + R$ we can take the molecular translational contribution to the molar heat capacity of a liquid as
$$
C_p^\text{tr}=20.78\,\text{J\,mol}^{-1}\text{K}^{-1}.
$$
Since the molar heat capacity of MIC is $111.76\,\text{J\,mol}^{-1}\text{K}^{-1}$ and that of water is $75.29\,\text{J\,mol}^{-1}\text{K}^{-1}$, on a molar basis the MIC-H$_2$O liquid system can store a large amount of heat in the internal rotational modes and in the vibrational intermolecular potentials. (The internal vibrational modes are not excited at this `slow' temperature.) 

The specific heat capacity of MIC, as opposed to the molar heat capacity,  is not particularly large, but that of water is anomalously high due to strong intermolecular forces. For the reaction mixture it is it is large enough to depress the first and third terms on the right hand side of equation \ref{e5} relative to the second term at low temperature. The rapid temperature spike and transition to `fast' time occurs because the activation energy is high enough to make the reaction very temperature-sensitive. In other words, the Arrhenius term in equation \ref{e5} can take over very rapidly  once the reaction zone reaches a certain temperature. At the temperature peak the system enters the 'fast' dissipative time interval, since the reactant is fully depleted. The specific heat capacity cannot affect the dynamics on this time scale. 

This analysis suggests that containment systems with large heat capacity for thermoreactive liquids may not suppress the oscillatory instability, although they would certainly lengthen the `slow' timescale. This may or may not be a good thing. A lengthy `slow' interval may provide enough time before the thermal runaway to deactivate or quench the system in other ways. On the other hand, it might give a false sense of security: \textit{`Nothing has happened over the last four hours, we might as well leave it.'}

\section{Conclusions\label{sec6}}

\begin{enumerate}
\item The CSTR paradigm was applied to investigate the thermal stability of MIC hydrolysis and the thermal decomposition of triacetone triperoxide in solution. 
\item Stability analyses of the steady state solutions of the dynamical model found that in both cases thermal runaway occurs due to the hard onset of a thermal oscillation at a subcritical Hopf bifurcation. Classical thermal ignition at a steady state turning point does not occur in these systems and over the thermal regime of interest they are dominated by oscillatory instability. 
\item This non-classical oscillatory thermal misbehaviour may be generic in liquid thermoreactive systems where the specific heat capacity and activation energy are high. 
These results provide new information  about the cause of thermal runaway that may inform improved designs of storage systems for  thermally unstable liquids and  better management of organic peroxide based explosives. 
\end{enumerate}

\appendix
\section*{Appendix}
The following brief account of the Bhopal disaster has been compiled from the following sources:
\citet{Forman:1985,Weir:1987,Shrivastava:1987,Lepkowski:1994} and \citet{Abbasi:2005}.  

The Union Carbide plant at Bhopal carried out the production of carbaryl, an agricultural insecticide that has been used widely throughout the world since 1945. Methyl isocyanate, a low-boiling point, highly reactive and extremely toxic liquid used in the synthesis of carbaryl, was stored in an underground stainless steel tank (Tank 610) which was encased in a concrete shell.  The temperature of the 41 tonnes of MIC in Tank 610 was 12--14$^\circ$C rather than the recommended 0--4$^\circ$C because the refrigeration unit had been non-operational for several months. On the evening of December 2 1984 a worker had been sent to hose out a nearby tank. The hose was left running unattended, and it is believed that a faulty valve allowed entry of water into the connected Tank 610. (Union Carbide disputes this, asserts that nothing was wrong with its equipment and procedures, and argues that sabotage by a disaffected employee must have led to the disaster.) By 11:30 pm, when workers detected lachrymose whiffs of leaking MIC, water had been running into Tank 610 for several hours. Although a slow rise in temperature and pressure in the tank had been noted, the early signs of trouble were not acted upon. Shortly after 11:30 pm the contents of the tank reached thermal criticality and began escaping as vapor from the flare tower.

Downwind of the flare tower lay the crowded suburbs and shantytowns. Most of the fluid in the tank streamed from the tower then drifted low over the city and sank and seeped in deathly mist in lungs in eyes, a period to sleep and swift arrest of retreat. About 4000 lives were claimed immediately, many tens of thousands through the subsequent days and months and years lost their lives or their health to the poison's effects, and the dead are still being counted. \\

\noindent\textit{Acknowledgement:}
The author is recipient of Australian Research Council Future Fellowship FT0991007.
\clearpage


\end{document}